\documentclass[12pt]{iopart}

\usepackage{hyperref}
\usepackage{amsfonts}
\usepackage{graphicx}
\usepackage{color,bm}

\begin{document}

\title[Beyond Bowen-York initial data]{Beyond the Bowen-York extrinsic curvature for spinning black holes}
\author{Mark Hannam, Sascha Husa, Bernd Br\"ugmann, Jos\'e A. Gonz\'alez, Ulrich Sperhake}
\address{Theoretical Physics Institute, University of Jena, 07743 Jena, Germany}

\date{\today}

\begin{abstract}
It is well-known that Bowen-York initial data contain spurious radiation. 
Although this ``junk'' radiation has been seen to be small for non-spinning 
black-hole binaries in circular orbit, its magnitude increases when the black 
holes are given spin. It is possible to reduce the spurious radiation by
applying the puncture approach to multiple Kerr black holes, as we
demonstrate for examples of head-on collisions of equal-mass black-hole binaries.
\end{abstract}


\maketitle

\section{Introduction}

Fully general relativistic numerical simulations of black-hole binaries
have one remarkable advantage over all analytic approximation techniques
(perturbation theory, close-limit analyses and post-Newtonian calculations):
given the correct initial 
data, simulations take into account {\it all} of the physics in the 
problem. One may be limited by computer resources and the accuracy of a given
numerical technique, but in principle no physics has been oversimplified or
overlooked in the calculation.  

We are left with one qualifier: the initial data must be physically correct. 
Producing good initial data involves finding physically realistic solutions of
the constraint equations of general relativity, and providing the appropriate
physical parameters for those solutions. The most commonly 
used initial data for recent evolutions of black-hole 
binaries have been puncture data based on the Bowen-York extrinsic curvature
\cite{Bowen80,Brandt97b} --- of the eight codes reported to have successfully
performed long-term evolutions of several orbits \cite{Pretorius:2005gq,
Campanelli:2005dd,Baker:2005vv,Herrmann:2006ks,Sperhake:2006cy,Scheel:2006gg,
Bruegmann:2006at,Pollney:NFNR}, six use Bowen-York puncture data.

One problem with these data is
that they contain unwanted spurious gravitational radiation. For example, a
spinning Bowen-York black hole is a Kerr black hole plus a Brill
wave \cite{Brandt94a}. In evolutions of black-hole binaries the spurious
radiation can be recognized as an initial ``bump'' in the extracted gravitational
waveform. It is interesting that this bump also appears in evolutions
performed using data produced within the conformal thin sandwich (CTS)
formulation of the initial-value equations \cite{Buonanno:2006n}, and is even
present in evolutions that start with boosted scalar fields that later
collapse to form black holes \cite{Pretorius:2005gq}.  

In Bowen-York data, we know that the initial burst of spurious
radiation is partly due to the Bowen-York form of the conformal extrinsic
curvature. The magnitude of the spurious radiation grows with both the linear
momentum and spin assigned to each black hole. 
We may therefore
worry that the ``junk'' radiation will significantly perturb the desired orbit
of the binary. For unequal-mass binaries this worry is justified: one sees
that the spurious radiation imparts a significant initial ``kick'' to the
black holes \cite{Gonzalez:2006md}. 

In this paper we describe a technique to reduce this spurious radiation for
spinning black holes, using a superposition of Kerr solutions in puncture form. 
This technique is similar in spirit to the superposition of Kerr-Schild black holes
\cite{Matzner98a,Marronetti00a,Marronetti00,Bonning:2003im}, but with
the advantage that it easily fits into the puncture framework that has proved
so useful for recent black-hole simulations. 
In the following sections we will summarize the construction of
black-hole initial data, outline the new method to produce superposed Kerr
punctures, and present first results of head-on collisions of Kerr
punctures, which demonstrate that the junk radiation has indeed been reduced.

\section{Decomposition of the initial-value equations}

We start with the standard 3+1 decomposition \cite{York79}, whereby the
Einstein equations are separated into initial-value and evolution
equations. An additional conformal decomposition results in initial-value
equations for which there are free data, which must be specified, and
constrained data that must satisfy the conformally decomposed constraint
equations.  

The Bowen-York solution to the momentum constraint was originally found in the
context of the conformal transverse-traceless (CTT) decomposition
\cite{Bowen80}. 
This solution has the great advantage that the linear momenta and spins of
multiple black holes can be easily specified; the global properties of the
data are free parameters. On the other hand, the choices of the free data in
the CTT decomposition necessary to obtain the Bowen-York solution were made
for convenience, and without strong physical motivation. It has been argued
\cite{Cook:2001wi} that better physically motivated choices can be made for
the free data in the more recent conformal thin-sandwich decomposition
\cite{York99}, and that these may lead to more astrophysically realistic
initial data.  

However, as we will see below, the most obvious physical choices of the free
data in the CTS decomposition are the {\it same} as those used to construct
the Bowen-York solution, and that the difference between the two constructions
may largely be a difference in coordinate choice. 

In the 3+1 decomposition of Einstein's equations, the constraints are coupled
elliptic equations for the spatial metric $\gamma_{ij}$ and extrinsic
curvature $K_{ij}$ on one time slice: 
\begin{eqnarray}
\nabla_j \left(K^{ij} - \gamma^{ij}K \right) & = & 0, \label{eqn:mc1} \\
R+K^2-K_{ij}K^{ij} & = & 0, \label{eqn:hc1}
\end{eqnarray} where $\nabla_i$ is the covariant derivative compatible with
 the spatial metric
$\gamma_{ij}$. The constraint equations can be further decomposed using a
conformal transformation. The spatial metric $\gamma_{ij}$ is related to a
background metric $\tilde{\gamma}_{ij}$ via a conformal factor $\psi$, 
\begin{equation}
\gamma_{ij} = \psi^4 \tilde{\gamma}_{ij}. \label{eqn:confg}
\end{equation} The extrinsic curvature is also decomposed. In the conformal
thin-sandwich decomposition \cite{York99}, we define $\tilde{u}_{ij} =
\partial_t \tilde{\gamma}_{ij}$. The extrinsic curvature is split into its
trace and tracefree parts, and can be written in terms of the
lapse function $N$ and the shift vector $\beta^i$, \begin{eqnarray}
K_{ij} & = & \psi^{-2} \tilde{A}_{ij} + \psi^4 \tilde{\gamma}_{ij} K \\
       & = & \psi^{-2} \left[ \frac{1}{2\tilde{N}} \left( (\tilde{\mathbb{L}} \beta)_{ij} 
       - \tilde{u}_{ij} \right) \right] + \psi^4 \tilde{\gamma}_{ij} K \\
       & = & \frac{1}{2N} \left[ (\mathbb{L} \beta)_{ij} - u_{ij} \right] +
       \gamma_{ij} K. \label{eqn:confKij}
\end{eqnarray} The conformal weights of the variables in these equations are 
\begin{eqnarray}
N & = & \psi^6 \tilde{N}, \\
\beta^i & = & \tilde{\beta}^i, \\
u_{ij} & = & \psi^4 \tilde{u}_{ij}, \\
K & = & \tilde{K}, \label{eqn:confK}
\end{eqnarray} 
and we are free to choose the function $\tilde{N}$. 

Let us now make the standard ``quasi-equilibrium'' choices of the
free data \cite{Cook:2001wi} that have been used when numerically producing
CTS data
\cite{Gourgoulhon02,Grandclement02,Cook:2001wi,Pfeiffer:2002xz,Pfeiffer:2002iy,Cook:2004kt,Ansorg:2005bp,Hannam:2005ed,Hannam:2005rp,Caudill:2006hw,Jaramillo:2006qj}:
conformal flatness, $\tilde{\gamma}_{ij} = f_{ij}$, maximal 
slicing, $K = 0$, and initial stationarity of the conformal metric,
$\tilde{u}_{ij} = 0$. If, in addition, we choose that $\tilde{N} = 1/2$, the
Hamiltonian and momentum constraints reduce to the same form as in the
conformal transverse-traceless decomposition: \begin{eqnarray}
\tilde{\nabla}_j \tilde{A}^{ij} & = & 0 \\
\tilde{\nabla}^2 \psi          & = & - \frac{1}{8} \psi^7 \tilde{A}_{ij}
\tilde{A}^{ij}.
\end{eqnarray}  The Bowen-York extrinsic
curvature is a solution to this form of the momentum constraint, and we see
that Bowen-York data satisfies all of the quasi-equilibrium requirements of
CTS data.  

The one extra ingredient in the standard construction of CTS data is another
quasi-equilibrium requirement, 
$\partial_t K = 0$, which yields an elliptic equation for $\tilde{N}$. This ``extended CTS''
system can be shown to admit non-unique solutions in some cases 
\cite{Pfeiffer:2005jf,Baumgarte:2006ug,Walsh:2006au}, but these problems 
have not prevented the construction
of initial data for black-hole binaries. The 
extra requirement $\partial_t K = 0$ will make some change to the physics of the system, 
but it is not clear what that change will be, and it is certainly unclear whether the
resulting data will be any more ``astrophysically realistic'' than Bowen-York
data. Initial-data studies that calculate the innermost stable circular orbit
(ISCO) for standard CTS data
\cite{Grandclement02,Cook:2004kt,Hannam:2005rp,Caudill:2006hw} find that its
location better matches post-Newtonian predictions than for Bowen-York data
\cite{Cook94,Baumgarte00a}. However, recent numerical simulations of
inspiralling black holes have cast doubt on how much sense it makes to discuss
an ISCO for comparable-mass black holes, so it is not clear how much weight to
give to the conclusions from these initial-data studies. One thing is certainly
clear from dynamic numerical simulations: moving from Bowen-York to CTS data
does not noticeably improve the spurious radiation content of the initial data. (Compare,
for example, Figure 2 in \cite{Baker:2006yw} with Figure~2 in \cite{Buonanno:2006n}.)

In the following, we will choose a function $\tilde{N}$ that is convenient for the 
numerical solution of the constraints. The solutions of the 
conformal thin-sandwich initial-value equations, together with the choice of $\tilde{N}$,
provide us with a lapse $N = \psi^6 \tilde{N}$ and a shift vector $\beta^i$, but
these are not necessarily ideal for dynamical evolutions. In particular, our choice for
$\tilde{N}$ will imply a lapse function that blows up at the punctures, and would produce
slices that hit the singularities in an infinitely short time! For evolutions we instead make
the standard choices $N = 1$ and $\beta^i = 0$ on the initial slice.

\section{Puncture approach}

Given the Bowen-York solution of the momentum constraint, we are left to solve
the Hamiltonian constraint to complete the specification of the initial data. It
is convenient to choose the ansatz \cite{Brandt97b}, \begin{equation}
\psi = 1 + \sum_i \frac{m_i}{2r_i} + u,
\end{equation} where the $m_i$ parametrize the masses of the black holes, and
the $r_i$ are the coordinate distances to each ``puncture''. The Hamiltonian
constraint now reduces to an 
elliptic equation for the function $u$. In this work, the Hamiltonian
constraint is solved using the fixed-mesh-refinement multigrid solver in the
BAM code \cite{Bruegmann:2006at}. 

There are many advantages to this approach. The $n$ black holes are
constructed with a topology that consists of $n+1$ asymptotically flat ends
connected by $n$ wormholes, and the slice, by construction, avoids the
singularities. As a result, the initial-value equations can be solved on all
of $R^3$, without any need to excise a region around each black hole.  A
second advantage is that it is possible to stably evolve these data, and
although connection with the extra asymptotically flat ends may be lost with
particular gauge choices \cite{Hannam:2006vv}, the slices remain singularity
avoiding, and one need never resort to excision techniques. 

When constructing Kerr-like punctures, we will also have to numerically solve
the momentum constraint. Although the conformal extrinsic curvature will
diverge at the punctures, we can ensure that the elliptic solver deals only
with non-divergent functions by a suitable choice of the conformal lapse,
$\tilde{N}$.

\section{Head-on collisions with spinning punctures}

Let us first consider the spurious radiation content of spinning Bowen-York
punctures. As a simple illustration, we will present evolutions of head-on
collisions of equal-mass black holes with their spins
directed away from each other. In this way, we expect the final black
hole to be non-spinning, i.e., a Schwarzschild black hole. 

Evolutions are performed with the BAM code \cite{Bruegmann:2006at},
 using the BSSN formulation of the evolution
equations \cite{Shibata95,Baumgarte99}. The lapse function and shift vector are
initially $N=1$ and $\beta^i = 0$, and are evolved using the 1+log \cite{Bona97a}
  and $\tilde{\Gamma}$-driver \cite{Alcubierre02a} slicing conditions. 

\begin{figure}[!ht]  
\centerline{\includegraphics[scale=0.55,angle=-90]{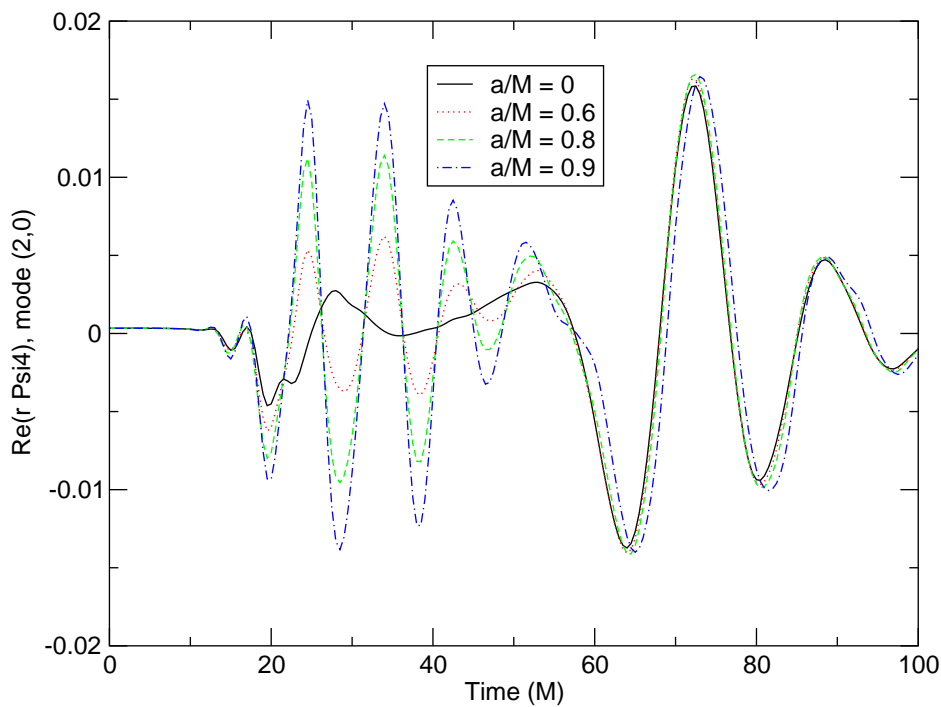}}
%\centerline{\includegraphics[scale=0.5]{fig1}}
\caption{\label{fig:BY_results}The spurious gravitational radiation from the
  head-on collision of equal-mass ($M_1 = M_2 = 0.5$) Bowen-York
  punctures. The punctures are placed at $z = \pm 4M$ and given spins $a/M_i =
  |S_z|/M_i^2 = 0,0.6,0.8,0.9$, such that the spins are directed away from the
origin.}
\end{figure}

Figure \ref{fig:BY_results} shows the results from runs that started with
punctures of masses $M_1 = M_2 = 0.5$ placed at $z = \pm 4M$ (where $M = M_1
+ M_2$), and given spins $|S_z| = 0,0.6,0.8,0.9M^2$, such that the spins were
directed away from the origin. The figure shows the ($l=2,m=0$) mode of the
real part of $r\Psi_4$, extracted at $r=20M$. (See \cite{Bruegmann:2006at} for a
description of the wave-extraction algorithm in the BAM code.) 

The gravitational radiation from the merger is clearly visible after about 60$M$
of evolution. Prior to that, some spurious radiation can be seen even when $S_z =
0$. This is presumably due to some unphysical component to the Brill-Lindquist
solution when the black holes are close. We do not expect the Brill-Lindquist solution
to be physically realistic unless the black holes are infinitely far apart, and indeed 
the spurious radiation decreases as the 
black holes are placed further apart, as is clearly seen in \cite{Sperhake:2006cy}. 
(See also \cite{Blanchet:2003kz} for a post-Newtonian extension of Brill-Lindquist
data.)

As spin is added, extra spurious initial radiation is seen. The extra spurious 
radiation has roughly the quasinormal mode frequency of each of the individual 
black holes. For the case where $|S_z| = 0.9M^2$, the initial ``junk'' radiation is
of comparable magnitude to the radiation from the black holes' merger. (We
should point out that this is an exaggeration over the more realistic case of
the merger of orbiting binaries, for which the merger waveform is far
stronger; see \cite{Campanelli:2006uy}.) 

\section{Kerr-like data for black-hole binaries}

Now we wish to reduce the initial radiation content for spinning
punctures. Motivated by Dain's work in \cite{Dain00,Dain:2001iw}, we
extend work done by Krivan and Price in axisymmetry \cite{Krivan:1998td}. We
start by writing the Kerr metric for a black hole with mass $m$ and angular
momentum $a/m$ in quasi-isotropic coordinates \cite{Brandt94a,Dain00}:
\begin{eqnarray} 
\tilde{\gamma}_{ij} & = & \delta_{ij} + a^2 h(m,a,r,\theta) \, v_{ij}, \\
v_{ij}              & = & \left\{(y^2,-xy,0),(-xy,x^2,0),(0,0,0)\right\}, \\
\psi               & = &  1 + \frac{\sqrt{m^2 - a^2}}{2r} + u_{\rm Kerr}
\label{eqn:KPsi} \\
\tilde{A}_{ij}      & = & \tilde{A}_{ij}^{\rm Kerr}.
\end{eqnarray} The analytic functional forms of $\tilde{A}_{ij}^{\rm Kerr}$ and
$u_{\rm Kerr}$ are given in \cite{Brandt94a} and \cite{Krivan:1998td}.

This form of the metric, which appears in \cite{Dain00}, can be generalized
to two black holes by superposition:  \begin{eqnarray}
\tilde{\gamma}_{ij} & = & \delta_{ij} + a_1^2 h_1 v_{ij}^1 + a_2^2 h_2
v_{ij}^2, \label{eqn:TwoKgamma} \\
\tilde{A}_{ij}     & = & \tilde{A}_{ij}^{K1} + \tilde{A}_{ij}^{K2} +
\frac{1}{2\tilde{N}} (\mathbb{L} \beta)_{ij}. 
\end{eqnarray} The background metric is a piece of the
free data, so a straightforward superposition is acceptable. The conformal
extrinsic curvature must satisfy the momentum constraint, so we consider it as
a superposition of the conformal extrinsic curvatures from two single Kerr
punctures, $\tilde{A}_{ij}^{K1}$ and $\tilde{A}_{ij}^{K2}$, plus a correction
that we must solve for numerically, and which we write in a form motivated by
(\ref{eqn:confKij}). We continue to make the choice of maximal slicing,
$K=0$. This construction is similar in spirit to that applied to Kerr-Schild
data by Matzner, Huq and Shoemaker \cite{Matzner98a}. Another related work
is \cite{Dain:2002ee}, where the authors adopted the Kerr form of the extrinsic 
curvature, but retained conformal flatness.

The constraints now take the form \begin{eqnarray}
\tilde{\nabla}^2 \psi & = & - \frac{1}{8} \psi^{-7} \tilde{A}_{ij} \tilde{A}^{ij} +
\frac{1}{8} \psi \tilde{R}, \\
\tilde{\Delta}_{\mathbb L} \beta^i - ( {\tilde{\mathbb L}} \beta )^{ij}
\tilde{\nabla}_j \ln \tilde{N} & = & - 2 \tilde{N} \tilde{\nabla}_j
(\tilde{A}^{ij}_{K1} + \tilde{A}^{ij}_{K2} ). \label{eqn:TwoKMC}
\end{eqnarray} Note that in the Kerr case the conformal Ricci scalar
$\tilde{R}$ is non-zero and diverges at the punctures. However, the Laplacian
of $\psi$ with respect to the background metric (\ref{eqn:TwoKgamma}) diverges
in the same way as $\psi \tilde{R}$, and the final right-hand side of the
Hamiltonian constraint is finite and sufficiently well-behaved that the code
is able to generate numerical solutions.

To solve the Hamiltonian constraint, we write the conformal factor in the
standard puncture form, \begin{equation}
\psi = 1 + \sum_i^n \frac{\sqrt{m_i^2 - a_i^2}}{2r_i} + u \equiv \psi_0 +
u,
\end{equation} where $\sqrt{m_i^2 - a_i^2}$ is motivated by the form of $\psi$
for the Kerr solution in quasi-isotropic coordinates, Eqn.~(\ref{eqn:KPsi}). 

In order to solve the momentum constraint numerically, we make a new choice
for the conformal lapse, \begin{equation}
\tilde{N}           =  \psi_0^{-3}
\end{equation} The reason for this choice is that the $\tilde{A}_{ij}^{K1,2}$
diverge at the punctures as $r^3$, but introducing a factor of $\psi^{-3} \sim
r^3$ makes the right-hand side of (\ref{eqn:TwoKMC}) finite at
the punctures. 

The constraints were solved using the multigrid solver in the BAM code
\cite{Bruegmann:2006at}, and the initial data $(\gamma_{ij},K_{ij})$
reconstructed using (\ref{eqn:confg}) -
(\ref{eqn:confK}). 

\begin{figure}[!ht]  
\centerline{\includegraphics[scale=0.5,angle=-90]{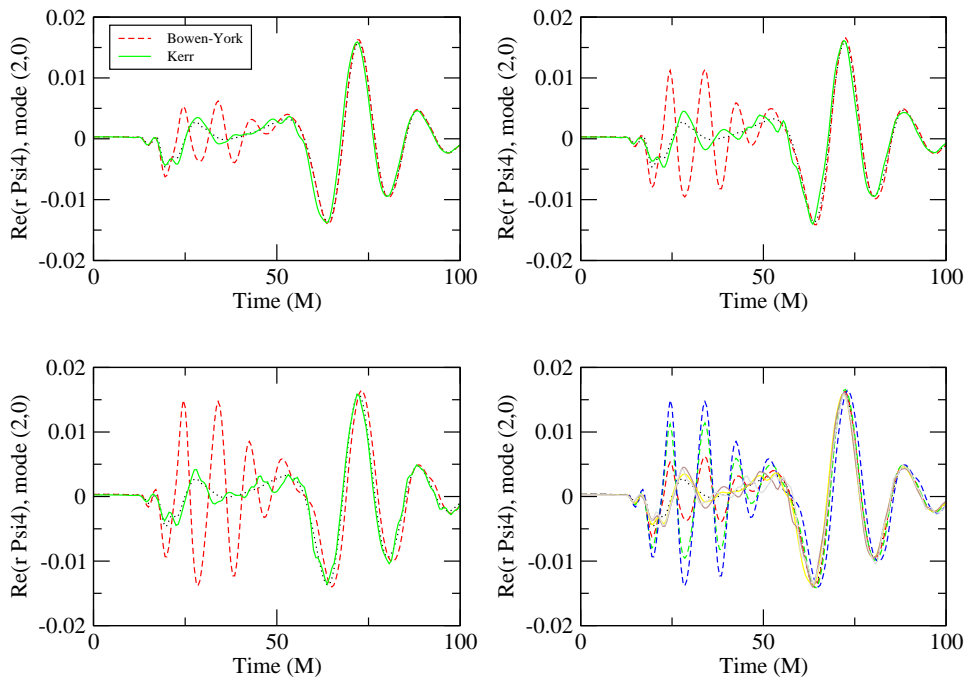}}
%\centerline{\includegraphics[scale=0.5]{fig2}}
\caption{\label{fig:kerr_results}Spurious radiation from head-on collisions of
  spinning Bowen-York and Kerr-like punctures, for spins of $a/M = 0.6$
  (top-left), $0.8$ (top-right), $0.9$ (lower-left), and all three
  (lower-right). The quantity shown in each plot is the $(l=2,m=0)$ mode of
  $r\Psi_4$ extracted at $20M$. In each plot the dotted black line corresponds
  to the radiation from a collision of nonspinning (Brill-Lindquist)
  punctures. In the top-left, top-right and lower-left plots, the dashed line
  shows the radiation from Bowen-York punctures, as also shown in
  Figure~\ref{fig:BY_results}, and the solid line shows the radiation from
  Kerr-like punctures. The lower-right plot allows a comparison between the
  three choices of spin.}
\end{figure}

Figure~\ref{fig:kerr_results} shows the results of evolving configurations
equivalent to those considered in Figure~\ref{fig:BY_results} for head-on
collisions of spinning Bowen-York punctures. We see that the spurious radiation
from the initial data deviates much less from that in the non-spinning
Brill-Lindquist case. There are some high-frequency components to the junk
radiation, but these may be due to the low tolerance used in the elliptic
solver; this requires more analytic and numerical study. The main result is 
clear: the superposed Kerr punctures contain dramatically less spurious 
radiation than their Bowen-York counterparts.

\section{Conclusions}

We have presented a method to construct initial data for two Kerr-like black
holes in the puncture framework. This construction is based on the form of the
Kerr metric in quasi-isotropic coordinates \cite{Brandt94a}, and following the
suggestions outlined in \cite{Dain00,Dain:2001iw} extends the work in axisymmetry
of \cite{Krivan:1998td}. 

For the simple test case of a head-on collision of two black holes, we have
demonstrated that the new initial data contain significantly less spurious
gravitational radiation than their popular Bowen-York counterparts. 

The results here serve as an illustration of the potential of this method. In
future work we will explore its effectiveness in more general configurations,
and extend the construction to spinning black holes that also possess linear
momentum. Such a construction would be necessary to make these data a feasible
alternative to Bowen-York data for simulations of black-hole binary inspiral.

\section*{Acknowledgments}

This work was supported in part by DFG grant SFB/Transregio~7 ``Gravitational Wave Astronomy''.
Computations where performed at HLRS (Stuttgart) and LRZ (Munich).

\section*{References}

\bibliographystyle{iopart-num}

\bibliography{references_cvs,references_extra}

\end{document}